\documentclass[longauth,letter,traditabstract]{aa}
\usepackage{graphicx}
\usepackage{wasysym}
\usepackage{natbib}
\usepackage{txfonts}
\begin{document}
   \title{{\it Herschel} measurements of the D/H and $^{16}$O/$^{18}$O ratios in water in the Oort-cloud comet C/2009 P1 (Garradd)\thanks{{\it Herschel} is an ESA space observatory with science instruments
  provided by European-led principal investigator consortia and with important contribution from NASA.}
  }

\author{D.~Bockel\'ee-Morvan\inst{1} \and N.~Biver\inst{1}
\and B.~Swinyard\inst{2, 3} \and M.~de Val-Borro\inst{4}  \and J.~Crovisier\inst{1} \and P.~Hartogh\inst{4}  \and D.C.~Lis\inst{5} \and R.~Moreno\inst{1} \and S.~Szutowicz\inst{6} \and E.~Lellouch\inst{1} \and M.~Emprechtinger\inst{5}  \and
G.A.~Blake\inst{5}  \and R. Courtin\inst{1}
  \and C.~Jarchow\inst{4}  \and M.~Kidger\inst{7} \and 
M.~K\"{u}ppers\inst{7}  \and M.~Rengel\inst{4} \and G.R.~Davis\inst{8}
  \and T. Fulton\inst{9} \and D. Naylor\inst{9} \and S. Sidher\inst{3} \and H. Walker\inst{3}  
 }


\institute{LESIA, Observatoire de Paris, CNRS, UPMC, Universit\'e Paris-Diderot, 5 place Jules Janssen, 92195 Meudon, France\\ 
\email{dominique.bockelee@obspm.fr}
\and Dept. of Physics \& Astronomy, University College London, Gower Street, London WC1E 6BT, UK                                                    
\and RAL Space, Science \& Technology Facilities Council, Rutherford Appleton Laboratory,
Chilton, Didcot, Oxon OX11 0QX, UK                        
\and Max Planck Institute for Solar System Research, Max-Planck-Str.~2, 37191 Katlenburg-Lindau, Germany           
\and California Institute of Technology, Pasadena, USA                         
\and Space Research Centre, Polish Academy of Science, Warszawa, Poland                  
\and European Space Astronomy Centre, Madrid, Spain                                     
\and Joint Astronomy Centre, 660 N. Aohoku Place, Hilo, HI 96720, USA                   
\and Institute for Space Imaging Science, Department of Physics and Astronomy,          
University of Lethbridge, Lethbridge, Alberta, Canada, T1K 3M4
        }

   \date{Received}


  \abstract{The D/H ratio in cometary water is believed to be an important indicator of the
conditions under which icy planetesimals formed and can provide
clues to the contribution of comets to the delivery of water and other
volatiles to Earth. Available measurements suggest that there is isotopic
diversity in the comet population. The {\it Herschel} Space
Observatory revealed an ocean-like ratio in the Jupiter-family
comet 103P/Hartley 2, whereas most values measured in Oort-cloud
comets are twice as high as  the ocean D/H ratio. We present here a new
measurement of the D/H ratio in the water of an Oort-cloud comet. HDO, H$_2$O, and $\mathrm{H}_2^{18}\mathrm{O}$ lines were
observed with high signal-to-noise ratio in comet C/2009 P1
(Garradd) using the {\it Herschel} HIFI instrument. Spectral maps of two water lines were
obtained to constrain the water excitation. The D/H ratio derived from the measured H$_2$$^{16}$O and HDO production rates is (2.06 $\pm$ 0.22) $\times$ 10$^{-4}$. This result 
shows that the D/H in the water of Oort-cloud comets is not as high as
previously thought, at least for a fraction of the population, hence the paradigm of a single, archetypal D/H ratio for all Oort-cloud comets is no longer tenable. Nevertheless, the value measured in 
C/2009 P1 (Garradd) is significantly higher than the Earth's ocean value of 1.558 $\times$ 10$^{-4}$. The
measured $^{16}$O/$^{18}$O ratio of 523 $\pm$ 32 is, however, 
consistent with the terrestrial value.}


\authorrunning{Bockel\'ee-Morvan et al.}
\titlerunning{D/H and $^{16}$O/$^{18}$O in the water of comet C/2009 P1 (Garrad)}
   \maketitle
%

\section{Introduction}
Having retained and preserved pristine material from the solar
nebula, comets contain unique
clues to the history and evolution of the solar system \citep{Irvine2000}.
Isotopic ratios are important indicators of the conditions under
which cometary materials formed, since isotopic fractionation is
very sensitive to physical conditions. In addition,
the characterization of the isotopic composition of long-period
comets from the Oort cloud and of short-period comets from
the Jupiter family can provide clues to their formation
regions in the early solar system.

The D/H ratio in water has been determined in several Oort-cloud comets (OCC)
 using different techniques, with most measurements agreeing
with a value of $\sim$ 3 $\times$ 10$^{-4}$  \citep[][and references therein]{Jehin09}. In
addition, the ratio was measured using the ESA {\it
Herschel} Space Observatory \citep{Pilbratt10} in the Jupiter-family comet (JFC) 103P/Hartley 2 \citep{hart11}, where it was found to
be (1.61 $\pm$ 0.24) $\times$ 10$^{-4}$, i.e., consistent with the
Vienna standard mean ocean
water (VSMOW) value equal to 1.558 $\times$ 10$^{-4}$ . This result was unexpected. Models
quantifying the deuterium enrichment factor in the solar nebula
with respect to the protosolar value predicted an increase in the
D/H ratio with increasing heliocentric distance
 \citep[e.g.,][]{Hersant2001,Kavel2011}. According to the most
accepted theory, OCCs and JFCs originate from the same population of
objects formed in the Uranus-Neptune zone \citep{Dones2004}, though part of the Oort cloud was possibly formed by planetesimals scattered from the Jupiter-Saturn region \citep{Brasser2008}. Therefore, JFCs were expected to exhibit a D/H ratio similar to or higher than that in OCCs \citep{Kavel2011}.

We present in this paper a new measurement of the D/H ratio in the
water of an Oort-cloud comet, obtained with the same
instrumentation and methodology as those used for comet 103P/Hartley 2. Comet
C/2009 P1 (Garradd) was observed with the Heterodyne Instrument for the Far-Infrared
\citep[HIFI,][]{2010HIFI}  in the framework of
the {\it Herschel} guaranteed time key programme ``Water and related chemistry in
the solar system'' \citep{hart09}. 
\vspace{-3mm}

\begin{table*}[t]
\setlength{\tabcolsep}{1.7mm}
\caption{\label{t1} HIFI observations of C/2009 P1 (Garradd) on 6
October 2011. Spectra characteristics and production rates.}
\begin{tabular}{lcccccc  l }
\hline\hline\noalign{\smallskip}
 & Line & $\nu$ & Mode & Date$^a$  & Line area$^b$ & $\Delta v$ & \phantom{000000000}Production rate$^i$  \\
& &&&&&& ~~~~$T_\mathrm{law}$~~~~~~~~~~~$T_\mathrm{kin}$ = 25 K~~~~~~$T_\mathrm{kin}$ = 47 K \\
&& (GHz) && (Oct. UT) & (mK km s$^{-1}$) & (m s$^{-1}$) & \phantom{00000000000000}(s$^{-1}$) \\
 \hline\noalign{\smallskip}
HDO & 1$_{10}$--1$_{01}$&509.29242 & Single-point & 6.428$^c$ & \phantom{00}22.5 $\pm$ 2.0\phantom{$^h$00} & --59 $\pm$ 33\phantom{$^h$} & $9.8(\pm0.9)\phantom{00}\phantom{//}9.2(\pm0.9)\phantom{00}\phantom{//}7.7(\pm0.6)\phantom{00}$ $\times$ $10^{25}$ \\
$\mathrm{H}_2^{18}\mathrm{O}$ & 1$_{10}$--1$_{01}$&547.67644 & Single-point & 6.415$^d$ & \phantom{0}179 $\pm$ 5\phantom{$^h$00} & --21 $\pm$ 13\phantom{$^h$} & $4.55(\pm0.12)\phantom{//}4.97(\pm0.12)\phantom{//}3.94(\pm0.11)$  $\times$ $10^{26}$\\
H$_2$O & 1$_{10}$--1$_{01}$& 556.93600  & Single-point & 6.415$^d$ & 9318 $\pm$ 12\phantom{$^h$0} & 320 $\pm$ 2\phantom{0$^h$} & $23.8(\pm0.6)\phantom{0}\phantom{//}23.4(\pm0.6)\phantom{0}\phantom{//}16.5(\pm0.4)\phantom{0}$ $\times$ $10^{28}$  \\
H$_2$O & 1$_{10}$--1$_{01}$& 556.93600 & Mapping & 6.291$^{e}$ & 8897 $\pm$ 98$^h$\phantom{0} & 331 $\pm$ 20$^h$ & $23.2(\pm1.6)\phantom{0}\phantom{//}26.2(\pm4.0)\phantom{0}\phantom{//}20.8(\pm5.5)$\phantom{0} $\times$ $10^{28}$  \\
H$_2$O & 1$_{10}$--1$_{01}$& 556.93600 & Mapping & 6.559$^{f}$ & 8109 $\pm$ 82$^h$\phantom{0} & 293 $\pm$ 16$^h$ & $21.4(\pm1.8)\phantom{0}\phantom{//}24.1(\pm3.7)\phantom{0}\phantom{//}19.2(\pm5.0)$\phantom{0} $\times$ $10^{28}$  \\
H$_2$O & 2$_{02}$--1$_{11}$&987.92676 & Mapping & 6.597$^{g}$ & 7140 $\pm$ 198$^h$  & \phantom{00}21 $\pm$ 19$^h$\phantom{0}& $18.2(\pm1.0)\phantom{0}\phantom{//}20.6(\pm2.1)\phantom{0}\phantom{//}15.5(\pm2.8)$\phantom{0} $\times$ $10^{28}$ \\

\noalign{\smallskip}\hline
\end{tabular}
\tablefoot{
\tablefoottext{a}{Mean date.} \tablefoottext{b}{Average of HRS and WBS line area
retrievals in main-beam brightness-temperature scale $T_{\rm mB}$. The error
bar corresponds to statistical noise, which is multiplied by 1.5 for single-point observations observed in FSW mode  to account for uncertainties in the baseline removal.} \tablefoottext{c}{Herschel observation identification number (Obsid)  1342230\#, with \# = 187, 189, 191, 193, 195, 197, 199, 201, 203, and 205.}
\tablefoottext{d}{Obsid: \# = 186, 188, 190, 192, 194, 196, 198, 200,
202, and 204.} \tablefoottext{e}{Obsid: \#185. } \tablefoottext{f}{Obsid: \#206.
} \tablefoottext{g}{Obsid: \#209.} \tablefoottext{h}{From the average of spectra
within 10\arcsec~of the peak brightness.} \tablefoottext{i}{For single-point data, a 4\arcsec~beam offset is considered (Appendix A). For mapping
data, the production rate corresponds to the weighted mean of
apparent production rates measured over the whole map, and the error bar is the dispersion
around the mean. }
}
\vspace{-3mm}
\end{table*}

\begin{figure}
\includegraphics[width=8.cm]{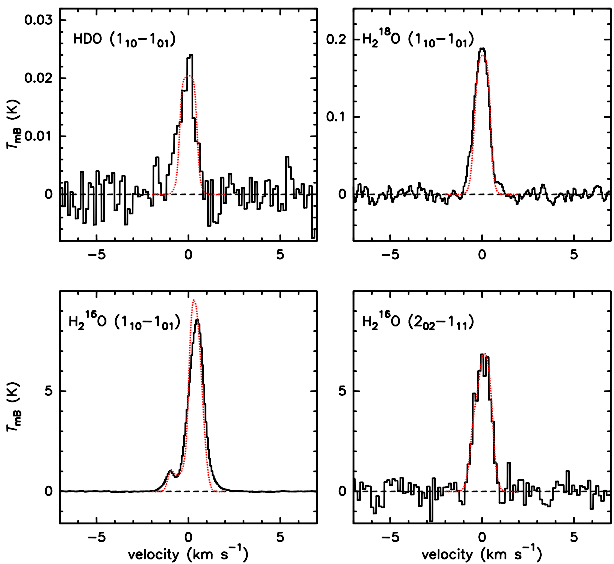}
\caption{HIFI spectra of comet C/2009 P1 (Garradd) observed on 6
October 2011 with the HRS. HDO (509 GHz), H$_2$O (557 GHz), and
$\mathrm{H}_2^{18}\mathrm{O}$ (548 GHz) 1$_{10}$--1$_{01}$ spectra are the average
of single-point measurements. The spectrum of the H$_2$O
2$_{02}$--1$_{11}$ line at 988 GHz is extracted from the map,
by averaging data at offsets $<$ 10\arcsec~from the peak.
The velocity scale is given with respect to the comet rest frame. Synthetic line 
profiles obtained with 30\% extended production (see text) are shown by red dotted lines.
Gas acceleration (velocity increasing from 0.48 km s$^{-1}$ to 0.58 km s$^{-1}$ from $r$ = 10$^4$ km to 10$^5$ km) is considered to fit more closely the wings of the profiles  \citep{combi2004}. }
\vspace{-5mm}
\label{fig:spectres}
\end{figure}

\section{Observations}

 Comet C/2009 P1 (Garradd) is a long-period comet originating from the Oort cloud \citep[$P$ = 127\,000
yr, orbit inclination of
106$^{\circ}$ with respect to the ecliptic;][]{Nakano}.  Discovered on 13 August 2009 at a heliocentric distance $r_h$ = 8.7 AU  \citep{2009IAUC.9062}, it passed
perihelion on 23 December 2011 at $r_h$ = 1.55 AU. The HDO observations 
with the HIFI instrument were performed on 6
October 2011, when the comet was at $r_h$ = 1.88
AU, and a distance from {\it Herschel} of 1.76 AU. The $\mathrm{H}_2^{18}\mathrm{O}$ and
H$_2$O lines were observed simultaneously. Since the H$_2$O
ground state rotational lines in comets are optically thick
\citep{Bensch2004,zakharov2007}, lines of the rare
oxygen isotopic counterpart $\mathrm{H}_2^{18}\mathrm{O}$ should, in principle, 
provide a more reliable reference for the D/H determination. 

The observing sequence followed the same scheme as that used for comet
103P/Hartley 2 \citep{hart11}. It consisted of ten 30-min long
observations of the HDO 1$_{10}$--1$_{01}$ rotational line at
509.292 GHz, interleaved with 6-min simultaneous  measurements of
the H$_2$O and $\mathrm{H}_2^{18}\mathrm{O}$ 1$_{10}$--1$_{01}$ ortho lines at
556.936 GHz and 547.676 GHz, respectively.  In
addition to these single-point measurements, two on-the-fly maps
of the H$_2$O 1$_{10}$--1$_{01}$ transition (of 16--min duration)
were acquired at the beginning and end of the sequence. Finally, a map of the  H$_2$O 2$_{02}$--1$_{11}$ para transition
at 987.926 GHz was performed with an integration time of
25 min. The full sequence spanned  6.85--14.71 UT on 6 October, with the maps serving to constrain the H$_2$O
excitation \citep{hart10,deval2010}.

The H$_2$O and $\mathrm{H}_2^{18}\mathrm{O}$ 1$_{10}$--1$_{01}$ lines were
observed in the upper and lower sidebands of the HIFI band 1a
mixer, respectively. The HDO line was also observed with the band 1a
mixer, whereas data on the H$_2$O 2$_{02}$--1$_{11}$ line at 988 GHz were 
acquired using the band 4a mixer.   The single-point observations were carried out in the frequency-switching
mode (FSW) with a frequency throw of 94.5 MHz. On-the-fly 557
GHz and 988 GHz maps were acquired using Nyquist sampling, and
 spatial coverages of 4\arcmin$\times$4\arcmin~and
2\arcmin$\times$2\arcmin, respectively. The observing mode for the
maps used a reference position at 10\arcmin~from the comet 
in RA. Spectra were acquired with both the Wideband
Spectrometer (WBS) and High Resolution Spectrometer (HRS). The
spectral resolution of the WBS is 1.1 MHz. The HRS was used either in
high-resolution (125 kHz) or nominal-resolution
mode (250 kHz), enabling us to sample the 
line shapes at a spectral resolution of 70--150 m s$^{-1}$. The
telescope beam sizes at the frequencies of the three lines
observed in band 1a are similar (half-power beam widths of
38\farcs1, 38\farcs7, and 41\farcs6 for the H$_2$O,
$\mathrm{H}_2^{18}\mathrm{O}$, and HDO lines, respectively), so that the three
molecules were observed over the same ($\sim$ 55\,000 km diameter) region
of the coma.

\begin{figure}[t]
\includegraphics[width=4.4cm, angle = 270, bb = 138 17 510 300]{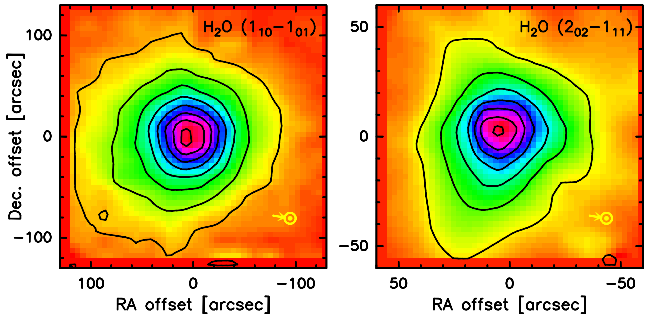}
 \caption{On-the-fly maps of H$_2$O in C/2009 P1 (Garradd) obtained with the WBS. Left: the $1_{10}$--1$_{01}$ 557 GHz line observed on 6.291 October 2011 UT.
 Right: the $2_{02}$--1$_{11}$ line at 988 GHz observed on 6.597 October 2011 UT. The contour spacing is 1 K km s$^{-1}$ in brightness temperature, corresponding to
 $\sim$ 6$\sigma$. The Sun direction is indicated at lower right. The beam sizes are 38\farcs1~and 21\farcs5~at 557 GHz and 988 GHz, respectively. } \label{fig:maps}
 \vspace{-5mm}
\end{figure}

Figure~\ref{fig:spectres} shows the HRS spectra of the single
pointing measurements (HDO, H$_2$O, and $\mathrm{H}_2^{18}\mathrm{O}$
1$_{10}$--1$_{01}$ lines), as well as the H$_2$O
2$_{02}$--1$_{11}$ spectrum extracted from the central part of the
map. Data reduction and calibration uncertainties are discussed in Appendix A.
The HDO line is detected with a line-integrated signal-to-noise ratio of 17.
Maps are shown in Fig~\ref{fig:maps}. Line intensities and
velocity offsets ($\Delta v$) in the comet rest-frame are given in
Table~\ref{t1}. The H$_2$O 1$_{10}$--1$_{01}$  line at 557 GHz is
optically thick and has an asymmetric profile owing to
self-absorption in the foreground coma. Intensity
variations with time of up to 7\% are observed for this line, which is 
related to intrinsic comet variability or instrumental effects.
Optically thin HDO and $\mathrm{H}_2^{18}\mathrm{O}$ lines have approximately
symmetric profiles. As the phase angle was 32$^{\circ}$, the small
negative-velocity offset observed for these lines suggests a modest excess
of outgassing toward the Sun.
\vspace{-3mm}

\section{Analysis}
The analysis was carried out using one-dimensional excitation
models of HDO, H$_2$O, and $\mathrm{H}_2^{18}\mathrm{O}$ \citep{biv07}. Models  
include collisions with H$_2$O and electrons, which dominate the
excitation in the inner coma, and solar infrared pumping of
vibrational bands followed by spontaneous decay, which establishes
fluorescence equilibrium in the outer coma. Radiation trapping
strongly affects H$_2$O excitation and is considered using the
escape probability formalism.  Consistent results were obtained using 
state-of-the-art radiation transfer methods \citep{hart10,Bensch2004,zakharov2007}. Synthetic
spectra are then computed using radiative transfer modelling. We assumed isotropic outgassing at a constant velocity. The model input parameters are: i) the gas expansion velocity, assumed
to be  0.6 km s$^{-1}$, corresponding to the half widths of optically thin 
HDO, H$_2$O  (988 GHz), and $\mathrm{H}_2^{18}\mathrm{O}$ lines; ii) the gas
temperature profile;  iii) the $x_{\rm ne}$ scaling factor of the
electron density profile, taken to be equal to 0.2
\citep{biv07,hart10,deval2010}. The H$_2$O ortho-to-para ratio is
assumed to be 3.

\begin{figure}
\includegraphics[width=6.0cm, angle = 270]{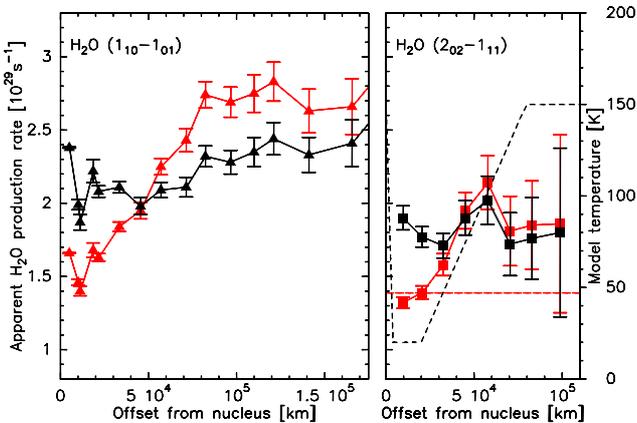}
 \caption{Apparent H$_2$O production rates as a function of beam offset deduced
 from the 557 GHz (left) and 988 GHz (right) H$_2$O maps observed on 6.56 and 6.60 October UT, respectively.
 Calculations with a 47 K and variable $T_\mathrm{kin}$  are shown with red and black symbols, respectively. The temperature profiles are plotted in the right panel with dashed lines. 
 } \label{fig:simul}
\vspace{-3mm}
\end{figure}

Line intensities for spectra probing the inner coma are sensitive
to the gas temperature profile, which controls both the population
of the rotational levels and the optical depth of the lines.
Hence, the evolution of the intensity of the H$_2$O
$1_{10}$--1$_{01}$ and $2_{02}$--1$_{11}$ lines with beam offset,
$\rho$ (km), carries information about the temperature profile.
Figure~\ref{fig:simul} presents the apparent H$_2$O production rate
$Q_\mathrm{app}$(H$_2$O) as a function of $\rho$ deduced from the maps. 
With appropriate modelling, the $Q_\mathrm{app}$ curve should be flat, if the nucleus is the dominant source of water vapour with a constant outgassing rate. Figure~\ref{fig:simul} presents the results for $T_\mathrm{kin}$ = 47 K,
a value consistent with multi-line observations of methanol undertaken
with the IRAM 30-m telescope in September and October 2011 with a 17\arcsec~beam comparable to the HIFI beam 
\citep{Biver2012}. For both lines, $Q_\mathrm{app}$(H$_2$O) increases with increasing $\rho$
for a constant temperature profile. Deviations from constant $Q_\mathrm{app}$(H$_2$O) are
enhanced when using $x_{\rm ne}$ $>$ 0.2 \citep{hart10}. The
temperature law $T_\mathrm{law}$ that minimizes the deviations of
$Q_\mathrm{app}$(H$_2$O) from a constant value, and provides consistent
(within 15\%) water production retrievals from the two lines, has
a minimum of 20 K at a distance $r$ = 4--20 $\times$ 10$^3$ km from the
nucleus  (thereby increasing the self-absorption and $Q_\mathrm{app}$ values) and then increases up to 150 K at $r$ = 8 $\times$ 10$^4$
km (Fig.~\ref{fig:simul}). This temperature increase may be related 
to the increased efficiency of photolytic heating \citep{combi2004}.  
We note that the line intensities are weakly sensitive
to the temperature at $r < $ 1000 km and $r > $ a few 10$^4$ km.
The velocity offsets $\Delta v$ of the on-nucleus synthetic
557(988) GHz line profiles are +292 (+50) and +259 (+39) m s$^{-1}$,
for $T_\mathrm{law}$ and $T_\mathrm{kin}$ = 47 K, respectively. 
Hence, the model with a variable temperature
also fits more closely the large positive $\Delta v$
of the central 557 GHz spectra (+320 m s$^{-1}$, Table~\ref{t1}),
as it enhances the optical thickness of the line.

We examined whether the increase in $Q_\mathrm{app}$(H$_2$O) with $\rho$
obtained with constant $T_\mathrm{kin}$ could instead be attributed to
water production from icy grains in the outer coma. For $T_\mathrm{kin}$ = 47 K, we are able 
to obtain a flat $Q_\mathrm{app}$(H$_2$O) profile for the 557 GHz line when assuming that
90\% of water production is extended with a characteristic Haser scale-length of $L_{\rm ext}$ = 30\,000 km. However this model is unsatisfactory since: i) the
apparent production rate given by the 988 GHz line now decreases
with increasing $\rho$; 2) the predicted $\Delta v$ of the 557 GHz
line is 30\% lower than observed; 3) the inferred $^{16}$O/$^{18}$O ratio in
water is then three times lower than the Earth value. On the other hand, subliming icy grains 
were possibly present in C/2009 P1 (Garradd)'s coma \citep{Paganini2012}. We therefore considered an alternative model with moderate (30\%) water production from long-lived icy grains (model outputs are  insensitive to an unresolved source of water). This model explains the water 557 and 987 GHz maps for $L_{\rm ext}$ = 50\,000 km and a temperature law similar to $T_\mathrm{law}$ (minimum of 20 K at 
$r$ = 4--16 $\times$ 10$^3$ km). The retrieved production rates and production rate ratios are identical 
(within 3\%) to those found for nuclear production with $T_\mathrm{law}$. This model accounts
satisfactorily for the observed line profiles (Fig.~\ref{fig:spectres}).

Table~\ref{t1} presents the production rates calculated with  
$T_\mathrm{law}$, along with those for constant $T_\mathrm{kin}$ = 25 K and 47 K. 
To compute the error bars in the production rate and isotopic ratios  (derived from the simultaneous single-point data), we 
take into account a 5\% relative calibration uncertainty (Appendix A). Using $T_\mathrm{law}$,
the HDO/$\mathrm{H}_2^{18}\mathrm{O}$ production rate ratio is 0.215 $\pm$ 0.023.
This is significantly larger than the value 0.161 $\pm$ 0.017 measured
for 103P/Hartley 2 \citep{hart11}. For $T_\mathrm{kin}$ = 25 K and 47 K,
one finds HDO/$\mathrm{H}_2^{18}\mathrm{O}$ = 0.185 and 0.195 ($\pm$0.020),
respectively, hence the retrievals are
only slightly temperature-dependent.

The H$_2^{16}$O/$\mathrm{H}_2^{18}\mathrm{O}$ production rate ratios derived from the band 1a observations using $T_\mathrm{kin}$ = 47 K is 419 $\pm$ 26, while for $T_\mathrm{law}$ and $T_\mathrm{kin}$ = 25 K one finds 523 $\pm$ 32 and 470 $\pm$ 29, respectively. These latter values are in good agreement with previous measurements in comets \citep[][ and references therein]{Jehin09}, as well as with the $^{16}$O/$^{18}$O = 498.7 VSMOW value, giving further confidence that our radiation-transfer model properly accounts for the opacity of the H$_2^{16}$O lines, provided a low gas temperature is adopted.

We adopt the model with $T_\mathrm{law}$ for the determination of
the D/H ratio, as it best explains the H$_2$O maps and line profiles.
The D/H measurement for 103P/Hartley 2 was based on the
HDO/$\mathrm{H}_2^{18}\mathrm{O}$ ratio \citep{hart11}, assuming $^{16}$O/$^{18}$O = 500 $\pm$ 50. Using the same method, we derive
D/H = (2.15 $\pm$ 0.32) $\times$ 10$^{-4}$ for C/2009 P1 (Garradd). 
Using instead directly the HDO/H$_2$$^{16}$O production-rate ratio 
results in D/H = (2.06 $\pm$ 0.22) $\times$ 10$^{-4}$. This value will be adopted for the discussion. We note that consistent results are derived using
constant (25--47 K) gas temperatures  and 30\% extended production (central values
ranging from 1.96 to 2.33 $\times$ 10$^{-4}$) .

 The water production rate derived using $T_\mathrm{law}$ is in the high range of values measured by other techniques in October 2011 (see discussion in Appendix B). Using a lower value would 
imply that the water of C/2009 P1 (Garradd) is highly enriched in $^{18}$O relative to both the Sun and all rocky bodies of the inner solar system \citep{McKeegan}. On the other hand, only an extreme  $^{18}$O enrichment 
 ($^{16}$O/$^{18}$O ratio of $\sim$ 330, i.e., $\delta^{18}$O = +500 \permil~in geochemical notation) would reconcile the D/H ratio in comet C/2009 P1 (Garradd) with the canonical Oort-cloud value of 3 $\times$ 10$^{-4}$.    
Such high $^{18}$O enrichments have not been found so far in any solar system body \citep{McKeegan}, except for CO in the Titan's atmosphere \citep{Courtin2011}. Our preferred interpretation is thus that the $^{16}$O/$^{18}$O ratio in comet Garradd is consistent with the VSMOW value.

\vspace{-5mm}

\section{Discussion}

\begin{figure}
\vspace{-2.5mm}
\includegraphics[width=5.2cm, angle = 270,bb = 245 57 510 180]{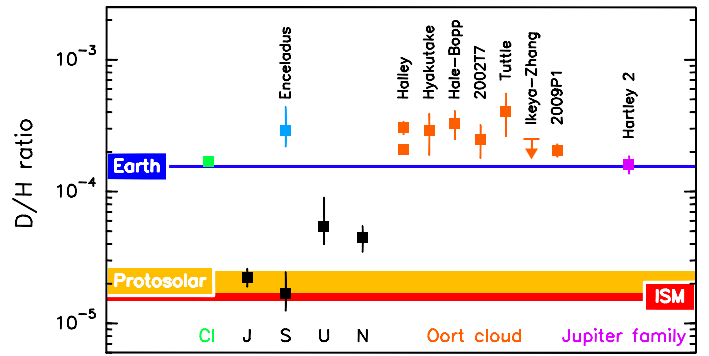}
 \caption{D/H ratio in the water of comets (including the revised value for Halley from \citet{Brown2012}) compared to values in carbonaceous meteorites (CI), the Earth's oceans, and Enceladus.  Displayed data for planets, interstellar medium (ISM), and the protosolar nebula refer to the value in H$_2$. Adapted from \citet{hart11}.} \label{fig:D/H}
 \vspace{-5mm}
\end{figure}

The discovery of a D/H value equal to that of the Earth's oceans in
the Jupiter-family comet 103P/Hartley 2 showed that the reservoir of
Earth-like water in the solar system is substantially larger than
previously thought, including now both carbonaceous meteorites and
comets \citep{hart11}. It also revealed that isotopic diversity is
present in the comet population, with members of the Oort cloud
having a deuterium enrichment of up to a factor of two with
respect to VSMOW (Fig.~\ref{fig:D/H}). As discussed by
\citet{hart11}, the suggested dichotomy between Oort-cloud and
Jupiter-family comets is difficult to explain in the context of
current models predicting deuterium enrichments in cometary ices,
and, if real, would imply to revisit the source regions of OCCs
and JFCs. Actually, the only dynamical theory that could explain
in principle an isotopic dichotomy in D/H is the one arguing  that
a substantial fraction of Oort-cloud comets were captured from
other stars when the Sun was in its birth cluster 
\citep{levison10}.

The paradigm for a single, archetypal D/H ratio for all Oort-cloud comets is no longer tenable. The value of (2.06
$\pm$ 0.22) $\times$ 10$^{-4}$ measured in comet C/2009 P1 (Garradd) is smaller
than the mean of previous determinations in Oort-cloud comets
\citep[(2.96 $\pm$ 0.25) $\times$ 10$^{-4}$, ][]{hart11},  and is consistent with the upper limit of 2.5 $\times$ 10$^{-4}$ measured in comet
153P/Ikeya-Zhang \citep[Fig.~\ref{fig:D/H},][]{Biv06}. We note that 
\citet{Brown2012}  reexamined mass-spectrometer
measurements for 1P/Halley \citep{eberhardt1995}, reevaluating this value
to be 2.1 $\times$ 10$^{-4}$. Altogether, the available
data suggest that the deuterium enrichment in the water of Oort-cloud comets is not as high as previously thought, at least for a
fraction of the population. Nevertheless, the D/H ratio measured in comet
C/2009 P1 (Garradd) is significantly higher (by 2.3 $\sigma$) than the VSMOW value
measured in 103P/Hartley 2. Interestingly, the range of D/H ratios measured in Stardust 
samples from comet 81P/Wild 2 \citep{stardust} is the same as measured in the bulk water of various comets.

Dynamical modelling suggests that the distribution of
planetesimals underwent large-scale mixing during the stages of
planetary migration \citep{Walsh2011}. Therefore, variations in
the cometary D/H may be expected within each population, if
ancestor reservoirs were isotopically different. Isotopic
diversity may be linked to the large compositional diversity
observed within both OCC and JFC populations \citep{bock2011}. 
Finally, experimental studies of ice sublimation suggest that the
D/H measured in the evaporated vapour might be enhanced or depleted
with respect to the bulk D/H in the cometary nucleus
\citep{Brown2012}. This demonstrates the need to increase the
sample of comets with accurate measurements of the D/H ratio,
as well as perform further modelling.

\begin{acknowledgements}
HIFI has been designed and built by a consortium of institutes and
university departments from across Europe, Canada and the United
States (NASA) under the leadership of SRON, Netherlands Institute
for Space Research, Groningen, The Netherlands, and with major
contributions from Germany, France and the US. Support for this
work was provided by NASA through an award issued by JPL/Caltech. 
SS was supported by polish MNiSW funds (181/N-HSO/2008/0).
\vspace{-4mm}

\end{acknowledgements}

\Online
\begin{appendix}
\section{Data reduction and HIFI calibration}
The comet was tracked using an up-to-date ephemeris provided by the
JPL Horizons system. The {\it Herschel} rms pointing accuracy is
approximately 1\arcsec.

The data were reduced to level 2 products using the {\it Herschel}
Interactive Processing Environment (HIPE 7.3). All lines were observed in the two
orthogonal H and V polarizations. The two orthogonal
polarizations were averaged. Note that the two polarizations are
observed with different mixers, and their respective apertures are
imperfectly co-aligned. The beam offset for the H and V average
spectra is $\sim$ 4\arcsec~with respect to the pointed position. 

The line intensities integrated over velocity were computed on the
main-beam brightness-temperature scale using beam efficiencies of
0.75 and 0.73 for bands 1a and 4a, respectively, and a forward
efficiency of 0.96. Based on the calibration error budget
\citep{Roelf2012}, a conservative value for the uncertainty in the
absolute intensity calibration is 10\% for both bands. Most
sources of errors are eliminated when comparing band 1a data, and
the relative uncertainty is at most 5\% in this case (sideband
ratio, hot-load coupling, and temperature). Finally, the
calibration uncertainty for the ratio of band 1a to band 4a lines
is 10\%.

\section{Water production rate: comparison with other measurements} 
Water production rates measured for C/2009 P1 (Garradd) from September to October 2011 are shown in Fig.~\ref{qh2o}. Reported values include retrievals from OH 18-cm observations (Colom et al. 2011, and in preparation), OH narrowband photometry (Schleicher et al., personal communication), and near-IR observations of water (DiSanti et al. 2012, Paganini et al. 2012; Villanueva et al. 2012). Herschel/HIFI and OH 18-cm observations, which were acquired  close in date, provide consistent values. The data suggest that the activity of comet Garradd underwent significant variations, and reached a maximum at the time the HIFI observations were performed. The 987 GHz H$_2$O line observed with HIFI was also observed on 16 October 2011 with the Spectral and Photometric Imaging Receiver (SPIRE) aboard Herschel; the line intensity is $\sim$ 40\% weaker than on 6 October 2011, when the HIFI measurements were conducted (Swinyard et al., in preparation), implying a water production rate consistent with the value measured from OH narrowband photometry on 18 and 20 October 2011 (65 days before perihelion, Fig.~\ref{qh2o}). The low values derived from the near-IR measurements, compared
to other measurements, possibly reflect sublimation from short-lived icy grains in the inner coma since the
field-of-view for these observations was much smaller (by a factor of ten or more) than for the other techniques.

\onlfig{5}{
\begin{figure}
\includegraphics[width=9cm]{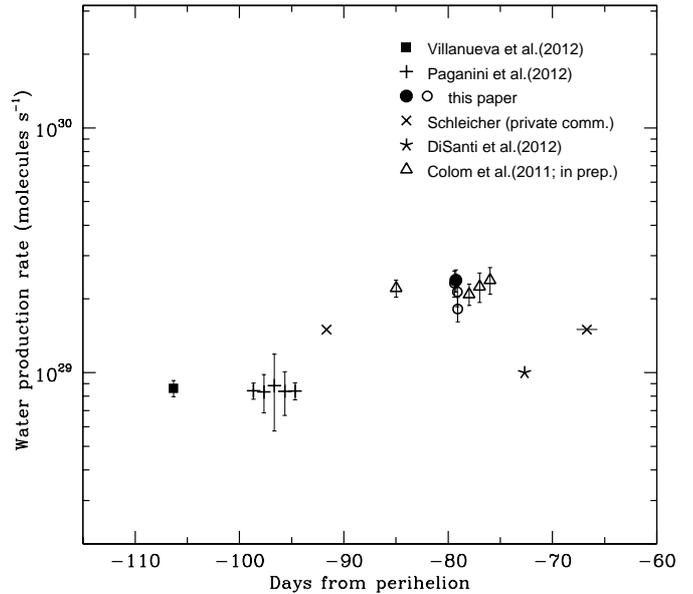}
\caption {Water production rates measured pre-perihelion in comet C/2009 P1 (Garradd). The time is with respect to the perihelion (23 December 2011). The plain and empty circles correspond to the 6-October single-point and mapping HIFI observations, respectively, analysed with the model with $T_\mathrm{law}$. } \label{qh2o}
\end{figure}
}

\end{appendix}

\end{document}